\documentclass[acmsmall]{acmart}
\AtBeginDocument{%
  \providecommand\BibTeX{{%
    Bib\TeX}}}

\usepackage{soul}
\usepackage{hyperref}
\usepackage{multirow}
\usepackage{listings}
\usepackage{fancybox}
\usepackage{enumitem}
\usepackage{graphicx}
\usepackage{wrapfig}
\usepackage{color}
\usepackage{xcolor}
\usepackage{array}
\usepackage{amsmath}
\usepackage{xspace}
\usepackage{url}
\usepackage{caption}
\usepackage{pgfplots}
\usepackage{balance}
\usepackage{subcaption}
\pgfplotsset{compat=1.16}
\usepackage{pgf-pie}
\usetikzlibrary{pgfplots.statistics,calc}
\usepackage{algorithm}
\usepackage{algorithmic}
\usepackage{adjustbox}
\usepackage{makecell}
\usepackage{tabularx}
\usepackage{longtable}
\usepackage{pdflscape}
\usepackage{tikz}
\usepackage{diagbox}
\usepackage{colortbl}

\usepackage{tcolorbox}
\makeatletter
\newcommand{\mybox}[1]{%
	\setbox0=\hbox{#1}%
	\setlength{\@tempdima}{\dimexpr\wd0+13pt}%
	\begin{tcolorbox}[boxrule=0.5pt, colback=white, arc=4pt,
		left=6pt,right=6pt,top=6pt,bottom=6pt,boxsep=0pt]
		#1
	\end{tcolorbox}
}

\definecolor{songcolor}{RGB}{191,191,191}

\usepackage{textcomp}
\usepackage{csquotes}

\def\BibTeX{{\rm B\kern-.05em{\sc i\kern-.025em b}\kern-.08em
    T\kern-.1667em\lower.7ex\hbox{E}\kern-.125emX}}

\usepackage{booktabs}
\usepackage{enumitem}

\usepackage{adjustbox}
\usepackage{multirow}

\newcommand{\tool}{\textit{SKILL-SIF}}

\definecolor{CardOrange}{HTML}{FFBA66}
\definecolor{CardYellow}{HTML}{FFF7E0}
\definecolor{CardBlue}{HTML}{E7F1FF}

\definecolor{DiffAddBG}{HTML}{E2F3E5}
\definecolor{DiffAddFG}{HTML}{1E7B34}
\definecolor{DiffDelBG}{HTML}{FCE2E2}
\definecolor{DiffDelFG}{HTML}{B3261E}

\newcommand{\cardfont}{\fontsize{6.2}{7.0}\selectfont}

\tcbset{
  pagentcard/.style={
    enhanced,
    boxrule=0.45pt,
    arc=1.4pt,
    left=0.4mm,right=0.4mm,top=0.8mm,bottom=0.8mm,
    coltitle=black,
    fonttitle=\scriptsize\bfseries,
    before upper={\cardfont\setlength{\parindent}{0pt}\raggedright},
  }
}

\newtcolorbox{observation}[1]{%
  enhanced,
  breakable,
  colback=gray!6,
  colframe=gray!60!black,
  boxrule=0.5pt,
  arc=2pt,
  left=6pt,right=6pt,top=4pt,bottom=4pt,
  title={\textbf{Observation:}~#1},
  fonttitle=\footnotesize\bfseries
}

\definecolor{sentpos}{RGB}{92,170,97}    
\definecolor{sentneu}{RGB}{220,220,220}  
\definecolor{sentneg}{RGB}{227,90,83}    

\newlength{\SentBarW}
\newlength{\SentBarH}
\newlength{\SentBarX}
\newlength{\SentGap}

\newlength{\SentPicW}
\newcounter{observation}
\newtcolorbox{observationbox}[1][]{
  colback=white,
  colframe=black!60,
  boxrule=0.6pt,
  arc=2pt,
  left=6pt,right=6pt,top=4pt,bottom=4pt,
  fonttitle=\bfseries,
  title={Observation~\stepcounter{observation}\theobservation},
  #1
}

\begin{document}

\title{How Effectively Do Security Skills Guide Coding Agents in Vulnerability Detection in Linux Kernel? 
}


 \author{Haoran Xue}
 \orcid{0009-0005-1937-3733}
 \affiliation{%
   \institution{York University}
   \department{Department of Electrical Engineering and Computer Science}
   \city{North York}
   \country{Canada}
 }
 \email{hrx00@yorku.ca}

\author{Nikta Akbarpour}
 \orcid{0009-0005-1937-3733}
 \affiliation{%
   \institution{York University}
   \department{Department of Electrical Engineering and Computer Science}
   \city{North York}
   \country{Canada}
 }
 \email{niktakbr@yorku.ca}

 \author{Gias Uddin}
 \orcid{0000-0003-1376-095X}
 \affiliation{%
   \institution{York University}
   \department{Department of Electrical Engineering and Computer Science}
   \city{North York}
   \country{Canada}
 }
 \email{guddin@yorku.ca}

 \author{Song Wang}
 \orcid{0000-0003-0617-2877}
 \affiliation{%
   \institution{York University}
   \department{Department of Electrical Engineering and Computer Science}
   \city{North York}
   \country{Canada}
 }
 \email{wangsong@yorku.ca}


\begin{abstract}
Large language model (LLM) coding agents, such as Claude Code, are increasingly used for security auditing, and task-specific skills have emerged as a practical way to provide them with reusable vulnerability-analysis procedures. However, it remains unclear how effectively existing security skills guide coding agents in identifying and localizing real vulnerabilities, how their effectiveness varies across vulnerability classes and workflows, and why skill-guided analysis fails. In this paper, we conduct the first large-scale empirical study of 37 security skills using Claude Code with four underlying LLMs on 100 real Linux kernel vulnerabilities spanning ten vulnerability classes, resulting in 14,800 skill-LLM-vulnerability runs. We evaluate each run on three complementary criteria: whether it identifies the correct vulnerable function, the responsible statement, and the vulnerability class. Across all runs, 24.4\% locate the correct function, 18.3\% identify both the function and responsible statement, and 15.7\% identify the correct vulnerable class. 
Our failure analysis further reveals recurring limitations in existing security skills, including insufficient support for evidence acquisition, state-relation reasoning, code interpretation, candidate verification, investigation coverage, and reporting.

Building on these findings, we develop {\tool}, an iterative self-improving framework that automatically evolves vulnerability detection skills by learning from previously missed vulnerabilities and their associated failure patterns. We apply {\tool} to five top-performing skills, using the original benchmark as the refinement set to evolve each skill and an independent set of 100 previously unseen vulnerabilities to evaluate the evolved skills. Our results show that {\tool} improves fully correct vulnerability detection by up to 8$\times$ (from 2\% to 16\%) on unseen vulnerabilities, with aggregate accuracy across Claude Code using four different underlying LLMs improving by 1.85$\times$ (from 6.05\% to 11.20\%). 
These results demonstrate that security skills can systematically learn from prior failures and transfer improved vulnerability-analysis strategies to unseen cases, providing practical insights for building more reliable skill-guided security agents. 
\end{abstract}




\maketitle

\section{Introduction}
\label{sec:intro}

Agent-based vulnerability detection is emerging as a promising approach to software security analysis~\cite{10.1145/3769082,JELODAR2026104390,ICLR2025_582d4e27}. 
Modern coding agents can autonomously explore repositories, inspect source code, invoke analysis tools, and reason about potential security weaknesses.
Their ability to combine repository-scale exploration with code understanding and natural-language reasoning makes them particularly attractive for vulnerability detection~\cite{10.1145/3769082,10.1145/3708522}.
Yet, detecting real-world vulnerabilities remains difficult because many vulnerabilities depend on subtle semantic relationships involving data flow, object lifetime, synchronization, resource management, validation, and execution state~\cite{10.1145/3806048}.

Agent skills have recently emerged as a practical way to provide coding agents with security-specific knowledge and analysis procedures~\cite{wang2025inducingprogrammaticskillsagentic,xu2026agentskillslargelanguage}.
A security skill can instruct an agent how to search for suspicious operations, trace relevant program states, validate vulnerability candidates, and collect supporting evidence, without modifying the underlying model. Public skill ecosystems already contain numerous security-oriented skills, while recent studies show that skills can substantially influence agent behavior and task performance~\cite{li2026skillsbenchbenchmarkingagentskills,
ling2026agentskillsdatadrivenanalysis}. Security skills are therefore becoming an important layer for agent-based vulnerability detection.

Despite their growing use, we still lack a systematic understanding of whether existing security skills actually guide coding agents to the correct vulnerabilities. A seemingly reasonable workflow may still narrow the search space too aggressively, omit critical state relationships, or allow candidates to be accepted or discarded without sufficient evidence. 
More importantly, vulnerability detection requires more than reaching approximately relevant code: a correct finding should identify the vulnerability class and localize the responsible code. Evaluating security skills therefore requires determining whether they guide agents to the actual root cause, rather than merely produce plausible vulnerability reports.

In this paper, we conduct a large-scale empirical study of public security skills for agent-based vulnerability detection. We collect security-related skills from three public skill marketplaces, systematically analyze their procedures, and select 37 representative skills covering diverse vulnerability detection workflows. We evaluate these skills using Claude Code with four underlying LLMs (i.e., Claude Sonnet-5, Claude Haiku-4.5, DeepSeek-V4-Pro, and DeepSeek-V3.2) on 100 real Linux kernel vulnerabilities spanning ten CWE classes, resulting in 14,800 controlled executions. For each execution, we examine whether the agent identifies the correct vulnerable function, reaches the responsible statement, and identifies the correct vulnerability class.

Our results show that existing skills provide useful but limited guidance.
Across all runs, 24.4\% locate the correct function, 18.3\% identify both the function and responsible statement, 15.7\% identify the correct CWE, and 11.1\% satisfy all three criteria. Performance varies substantially across vulnerability
classes, skills, and underlying LLMs. Bounds-related vulnerabilities are detected more reliably than vulnerabilities requiring lifetime, synchronization, resource, or execution-state reasoning, while higher-performing skills commonly combine broad candidate discovery with a subsequent reasoning or verification stage. Our failure analysis further identifies recurring breakdowns in evidence acquisition, state-relation reasoning, verification, candidate retention, and reporting.

These findings raise a further question: \textit{can security skills improve by learning from their own failures?} To investigate this possibility, we develop {\tool}, an iterative self-improving framework that refines vulnerability detection skills using previously missed vulnerabilities and their diagnosed failure patterns. 
We apply {\tool} to the five highest-performing skills and evaluate the refined skills on an independent set of 100 previously unseen vulnerabilities. The refined skills consistently improve vulnerability detection, with fully correct detection increasing by up to 8$\times$ and
aggregate accuracy across four LLMs used with Claude Code improving by
1.85$\times$ (from 6.05\% to 11.20\%). 
These results demonstrate that execution feedback can be used to systematically improve security skills and that the resulting improvements can generalize to unseen vulnerabilities.

 

In summary, our study provides the following contributions:\begin{itemize}[leftmargin=10pt]
    \item We conduct a large-scale empirical evaluation of 37 security skills using Claude Code with four underlying LLMs on 100 real Linux kernel vulnerabilities, producing 14,800 controlled executions. 
    \item We characterize how effectively existing security skills guide vulnerability detection and localization, and analyze how their effectiveness varies across vulnerability classes, skills, and underlying LLMs. 

    \item  We analyze unsuccessful execution trajectories and develop a taxonomy
    that distinguishes agent-execution failures from skill-level limitations.
    
    \item We develop {\tool}, an iterative framework that learns from failed
    executions to refine security skills. Evaluation on 100 previously unseen
    vulnerabilities shows improvements of up to 8$\times$, with aggregate
    fully correct detection improving by 1.85$\times$ (6.05\% to 11.20\%). 
\end{itemize} 

The remainder of this paper is organized as follows. Section~\ref{sec:related} reviews related work. Section~\ref{sec:empirical_study} presents the study design, and Section~\ref{sec:result} reports the empirical results. Section~\ref{sec: skill-sif} introduces {\tool} and evaluates in Section~\ref{sec: skill-sif-performance}. Section~\ref{sec:discussion} discusses implications, Section~\ref{sec:threats} discusses threats to validity, and Section~\ref{sec:conclusion} concludes.
\section{Related Work}
\label{sec:related}

\subsection{Vulnerability Detection: From Program Analysis to LLMs}
Vulnerability detection has traditionally relied on program analysis~\cite{avgustinov2016ql, 216013, 10.1145/3533767.3534380}, kernel fuzzing~\cite{217573, inproceedings, 274553}, symbolic execution~\cite{217563, 255310}, sanitizers~\cite{287286}, and task-specific learning models~\cite{Li_2018, Li_2022}. In the Linux kernel, Coccinelle uses semantic patches to detect and transform recurring code patterns~\cite{216013}, while fuzzing systems such as MoonShine~\cite{217573}, HFL~\cite{inproceedings}, and SyzVegas~\cite{274553} extend Syzkaller-style vulnerability discovery~\cite{syzkaller} through seed distillation, fuzzing, and learning-based task selection.

More recently, LLMs have been studied for vulnerability detection, classification, repair, and explanation. Pearce et al.~\cite{pearce2023zeroshot} examine zero-shot vulnerability repair and highlight both its potential and sensitivity to model behavior, while Ullah et al.~\cite{ullah2024llms} show that LLM-based security reasoning can be non-deterministic and sensitive to superficial changes. Other studies evaluate LLMs and code models for vulnerability detection and weakness classification~\cite{lin2025mammoth,ding2025primevul}. LLMs have also been combined with static analysis, including the triage and explanation of Semgrep findings~\cite{nguyen2025llmassisted} and the generation or assistance of CodeQL queries~\cite{wang2026cqllm}. Most of this work evaluates LLMs as standalone analyzers, classifiers, repair systems, or companions to existing security tools. Our study instead examines whether reusable procedural skills can guide a tool-using coding agent with different underlying LLMs to correctly attribute and localize real vulnerabilities in Linux kernel source code.

\subsection{Skills and Procedural Guidance for Coding Agents}

Agent skills provide reusable task-specific instructions, resources, scripts, and tool-use guidance that can be loaded at inference time without retraining the underlying model. Ling et al.~\cite{ling2026agentskillsdatadrivenanalysis} conduct a large-scale empirical analysis of public Claude-style skills and find substantial representation of software-engineering workflows, together with redundancy, uneven quality, and safety concerns. SkillsBench~\cite{li2026skillsbenchbenchmarkingagentskills} evaluates curated skills across diverse agentic tasks and shows that skills can improve agent performance, although the gains vary across tasks and domains. Other work studies the security risks introduced by skill packages. Liu et al.~\cite{liu2026agenticskillsworkwild} identify risks including prompt injection, data exfiltration, privilege escalation, and supply-chain exposure, while SkillInject~\cite{schmotz2026skillinjectmeasuringagentvulnerability, schmotz2025agentskillsenablenew} further demonstrates that skills can create new attack surfaces through injected instructions and unsafe tool use. Together, these studies establish skills as an important adaptation layer for agent behavior. Our study focuses on their effectiveness for security analysis, evaluating whether public vulnerability-detection skills can guide coding agents to correctly identify the vulnerability class and localize the responsible function and statement in real Linux kernel revisions.

\subsection{Skill Refinement and Evolution}
Recent work has also explored how agent skills can be improved from execution experience. SkillRevise~\cite{liu2026skillreviseimprovingllmauthoredagent} diagnoses defects from execution traces, retrieves relevant repair principles, revises the skill, and validates the result through re-execution. EvoSkills~\cite{zhang2026coevoskillsselfevolvingagentskills} extends skill evolution to multi-file packages using a co-evolving verifier, while Liu et al.~\cite{liu2026rethinkingselfevolvingagentskills} study how multi-round feedback and validation affect skill evolution. Trace2skill~\cite{ni2026trace2skilldistilltrajectorylocallessons} distills recurring lessons from execution trajectories into transferable skills, and SkillAdaptor~\cite{yu2026skilladaptorselfadaptingskillsllm} performs targeted skill updates through step-level failure attribution. SkillOpt~\cite{yang2026skilloptexecutivestrategyselfevolving} formulates skill refinement as controlled text-space optimization with roll-based validation, while Memskill~\cite{zhang2026memskilllearningevolvingmemory} evolves memory skills by learning from difficult cases. Yang et al.~\cite{yang2026defenseasskillevolvingruntimeguard} evolve a runtime guard skill from rollout feedback, while TRUSS~\cite{zhang2026trusstaskreliableusersafeautomated} uses execution failures and safety violations to iteratively refine generated skills. Related prompt-optimization work such as GEPA~\cite{agrawal2026gepareflectivepromptevolution} similarly uses execution trajectories and natural-language reflection to evolve procedural instructions without updating model parameters. These studies primarily develop general-purpose skill generation and refinement. Our study focuses specifically on security skills for source-code vulnerability detection, introducing a failure-mechanism-driven evolutionary refinement process that learns from missed real-world vulnerabilities.

\section{Empirical Study Design}
\label{sec:empirical_study}


\begin{figure*}[t!]
  \centering  \includegraphics[width=\textwidth]{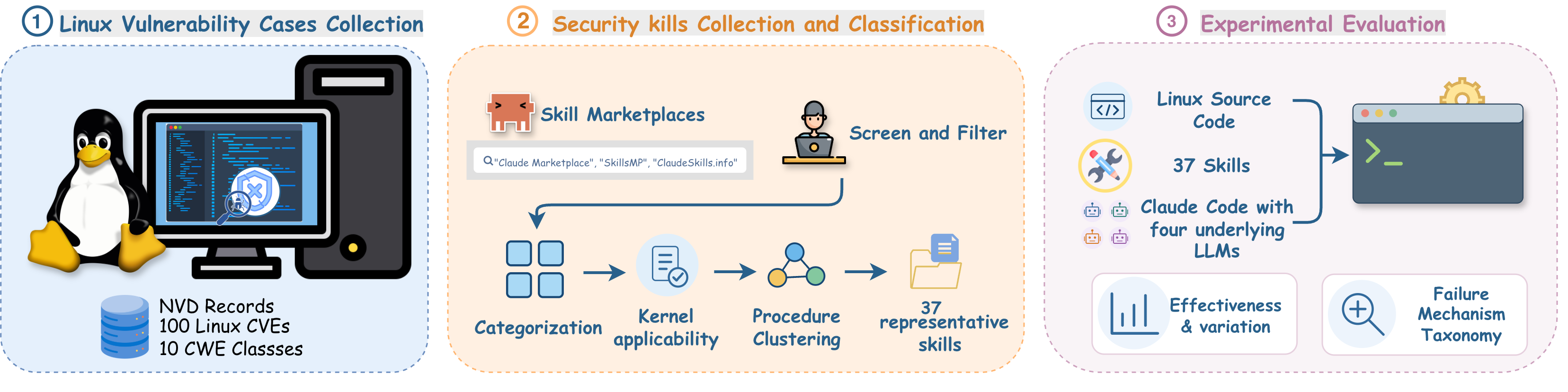}
   \vspace{-0.8cm}
  \caption{Empirical study overview} 
\label{fig:workflow}
\vspace{-0.2cm}

\end{figure*}
Figure~\ref{fig:workflow} summarizes the overall empirical study on evaluating security skills for vulnerability detection. We evaluate representative public skills on real Linux kernel vulnerabilities and examine their effectiveness, variation across vulnerability classes and workflow patterns, and recurring failure mechanisms. This empirical study is organized around RQ1–RQ3 and is reported in Section~\ref{sec:result}. 


\subsection{Research Questions}
\label{subsec:research-questions}

We organize the study around the following research questions.

\begin{itemize}[leftmargin=10pt]
    \item \textbf{RQ1 (Effectiveness):} How effectively do existing security skills guide a model to detect and localize real Linux kernel vulnerabilities? We measure whether a run identifies the correct vulnerability class and localizes the responsible function and statement.

    \item \textbf{RQ2 (Performance variation):} How does effectiveness vary across skill vulnerability classes and skills with different workflow patterns? We compare detection effectiveness across vulnerability classes and across skills with different procedure workflows.

    \item \textbf{RQ3 (Failure mechanisms):} What failure mechanisms prevent skill-guided runs from correctly detecting and localizing the vulnerability? We inspect the execution trajectory to characterize failure patterns and develop a taxonomy.


\end{itemize}


\subsection{Case and ground truth construction}
\label{subsec:data-prepare}
\begin{table}[t]
\centering
\caption{Top-10 CWE Types in the Dataset}
\label{tab:top-10-cwes}
\footnotesize
\setlength{\tabcolsep}{3pt}
\renewcommand{\arraystretch}{0.7}

\begin{tabular}{@{}cll@{}}
\toprule
\textbf{Rank} & \textbf{CWE} & \textbf{Weakness Type} \\
\midrule
1  & CWE-416 & Use After Free \\
2  & CWE-476 & Null Pointer Dereference \\
3  & CWE-401 & Missing Release of Memory after Effective Lifetime \\
4  & CWE-125 & Out-of-bounds Read \\
5  & CWE-787 & Out-of-bounds Write \\
6  & CWE-667 & Improper Locking \\
7  & CWE-908 & Use of Uninitialized Resource \\
8  & CWE-825 & Expired Pointer Dereference \\
9  & CWE-362 & Concurrent Execution using Shared Resource with Improper Synchronization \\
10 & CWE-617 & Reachable Assertion \\
\bottomrule
\end{tabular}
\end{table}
We select the Linux kernel as our experimental domain because it is a large, mature, and security-critical system that exhibits a wide range of real-world vulnerability categories and mechanisms~\cite{kroahhartman2007linuxkerneldevelopment, 10.1145/3729423}. We construct a source collection of Linux kernel vulnerabilities published by the National Vulnerability Database (NVD)\footnote{https://nvd.nist.gov/} during June 2026, from June 1 inclusive to July 1 exclusive. We retain a Common Vulnerabilities and Exposures (CVEs) record if it identifies the Linux kernel as affected and provides a public fixing-commit reference with an identifiable parent revision, a recoverable patch, and at least one source file modified by the fix that exists in the parent revision. 
We rank Common Weakness Enumeration (CWE) classes by the number of distinct Common Vulnerabilities and Exposures (CVE) records assigned to each class. 
The ten highest-frequency classes are listed in the Table~\ref{tab:top-10-cwes}, which are CWE-416, CWE-476, CWE-401, CWE-125, CWE-787, CWE-667, CWE-908, CWE-825, CWE-362, and CWE-617. These classes define the weakness coverage of our benchmark. For each selected class, we choose the ten most recently published eligible CVEs before July 1, 2026. A CVE associated with multiple selected classes is included only once, and the remaining class is completed using the next most recent eligible CVE. This procedure produces 100 distinct CVE-level cases, with exactly ten cases per CWE. 

Each case represents one CVE and one before-fix repository snapshot, constructed from the first parent of the selected fixing commit. All source files modified by that fix are retained as the case's target paths. A fix modifying multiple files therefore remains a single evaluation case rather than being decomposed into separate file-level cases. Correspondingly, a case may contain multiple valid vulnerable functions and responsible statements across its target files. The complete before-fix repository snapshot is preserved so that related code outside the target files remains available as contextual information during execution. For each case, we construct ground-truth annotations by comparing the fixing patch with the corresponding before-fix and fixed source code. We distinguish operations directly responsible for the vulnerability from supporting changes such as declarations, refactoring, comments, or diagnostic updates. When a patch introduces a previously missing validation, cleanup, or other safety operation, we trace that change to the corresponding unchecked operation or execution path in the before-fix revision. Each responsible statement is annotated with its source text, enclosing function where applicable, and before-fix line range. Statements spanning multiple source lines are treated as a single program operation. A localization is considered a statement match when the reported code or line range unambiguously identifies an annotated responsible operation; listing every continuation line is not required. Function and statement correctness are assessed against all annotated valid locations for the case, allowing multiple correct functions or statements when the vulnerability spans more than one location. 

\subsection{Skill Collection, Classification, and Selection}
\label{subsec: skill-select}

We construct the skill set through a multi-stage collection and filtering process that progressively narrows public security skills to a procedurally diverse set suitable for Linux-kernel vulnerability detection. We first collect listings from three major agent-skill marketplaces: Claude Marketplace\footnote{https://claudemarketplaces.com/}, SkillsMP\footnote{https://skillsmp.com/}, and ClaudeSkills.info\footnote{https://claudeskills.info/}, and verify that each candidate maps to an accessible GitHub repository containing a valid SKILL.md, and remove duplicates, which yields 12,402 distinct skills. 
We then use Claude Opus-5 to inspect each skill's complete SKILL.md and relevant repository content and determine whether vulnerability detection is a substantive part of its procedure. Skills whose primary role is vulnerability discovery, assessment, or triage are retained, while skills limited to remediation, generic code review, reference material, or unrelated security operations are excluded. We cross-check the screening decisions with a second model (gpt-oss-20B) and manually resolve disagreement cases, resulting in 7,476 vulnerability-detection skills. We next conduct two distinct classifications of these skills: Primary Functional category (PFC) and Methods/Technique category (MTC), based on the actual skill instructions and repository content. The primary functional category indicates the main purpose of the skill, while the methods/techniques category indicates the primary analysis or execution method used by the skill. These two categories provide us with a clearer overview and understanding of the skills from different aspects. The complete PFC and MTC categories and distributions are reported in the supplementary material; Table~\ref{tab:pfc_mtc_filtering} summarizes the categories used in the subsequent Linux-kernel filtering.  

\begin{table*}[t]
\centering
\caption{PFC and MTC dimensions used for Linux-kernel skill filtering.}
\label{tab:pfc_mtc_filtering}

\small
\setlength{\tabcolsep}{3pt}
\renewcommand{\arraystretch}{0.8}

\begin{tabularx}{\textwidth}{
    @{}
    >{\centering\arraybackslash}p{0.10\textwidth}
    >{\raggedright\arraybackslash}p{0.42\textwidth}
    >{\raggedright\arraybackslash}X
    @{}
}
\toprule
\textbf{Dimension} &
\textbf{Classification question} &
\textbf{Categories used for kernel filtering} \\
\midrule

\textbf{PFC} &
What is the skill primarily intended to do? &
PFC-01: Source Code Vulnerability Detection \\

\textbf{MTC} &
What is the skill's primary analysis or execution method? &
MTC-01: LLM-Native Static Reasoning; 
MTC-02: External Security-Tool Orchestration; 
MTC-03: Pattern/Regex/AST Grep Scanning; 
MTC-06: Fuzzing \& Property-Based Testing; 
MTC-12: Bundled Custom Analyzer Scripts \\

\bottomrule
\end{tabularx}
\vspace{-0.3cm}
\end{table*}

We next restrict the corpus to source-code vulnerability-detection skills whose analysis methods can operate on Linux-kernel source or source-derived artifacts. This produces 1,460 candidates for detailed applicability review. Based on the procedures and dependencies described by each skill, we identify 33 directly applicable and 367 generally applicable skills, and retain these 400 skills for procedure-level analysis. For each skill, we characterize how it selects suspicious code, explores related program state and context, verifies candidate vulnerabilities, and preserves evidence in the final finding. Skills with substantively equivalent
investigation procedures are grouped and manually consolidated into 50 procedure clusters. We select one available skill from each cluster that claims coverage of at least one benchmark CWE; when multiple candidates are available, we use the one with the highest recorded GitHub-star count. This yields 37 representative skills from 37 distinct procedure clusters. The clusters are used to ensure procedural diversity; all reported performance corresponds to the selected individual skills rather than to their entire clusters.

\subsection{Coding Agent and Underlying LLMs}
\begin{table*}[t]
\centering
\caption{Evaluated LLMs, versions, and release dates.}
\label{tab:model}

\footnotesize
\setlength{\tabcolsep}{6pt}
\renewcommand{\arraystretch}{0.9}

\begin{tabular}{@{}llll@{}}
\toprule
\textbf{Model} &
\textbf{Current Version} &
\textbf{Release Date} &
\textbf{Ref.} \\
\midrule

Claude Sonnet-5 &
Claude Sonnet-5 &
June 30, 2026 &
\cite{anthropic_transparency} \\

Claude Haiku-4.5 &
Claude Haiku-4.5 &
Oct. 15, 2025 &
\cite{claude_haiku45} \\

DeepSeek-V4-Pro &
v4-pro-0813 &
Aug. 13, 2026 &
\cite{deepseek_updates} \\

DeepSeek-V3.2 &
DeepSeek-V3.2 &
Dec. 1, 2025 &
\cite{deepseek_v32} \\

\bottomrule
\end{tabular}

\end{table*}

We evaluate the selected skills with four underlying LLMs spanning two model families and different capability levels, as presented in Table~\ref{tab:model}. We include two Claude models and two open-source models from DeepSeek to reduce dependency on a single model family. Within each family, we select models representing different capability tiers: Claude Haiku-4.5 provides a lighter-weight model while Claude Sonnet-5 represents a stronger model, and DeepSeek-V3.2 and DeepSeek-V4-Pro provide a similar contrast across two generations of the DeepSeek family. This choice allows us to compare the skill's effectiveness across different LLMs.

\subsection{Experimental Setup}
\label{subsec:experimental-setup}
We evaluate each selected skill through the Claude Code command-line interface in non-interactive mode. Each skill-case pair runs in a fresh workspace containing the complete before-fix Linux source tree without Git history and the selected skill package. The model is provided with the case's target file paths and asked to inspect them for security vulnerabilities, while allowing it to examine other files in the supplied repository for contextual reasoning. Our evaluation therefore focuses on file-guided vulnerability identification and localization. All runs follow the same review scope and access restrictions, while skill-specific instructions retain each skill's analysis workflow and reporting format. We configure explicit tool permissions and instruct the agent to use only the supplied repository and selected skill, without accessing internet resources, external files, Git history, or vulnerability databases. For every run, we record the execution trajectory, run metadata, generated report artifacts, and final response. We extract the reported vulnerability claims and code locations into a common evaluation format to compare them with the ground-truth CWE, function, and statement annotations defined in Section~\ref{subsec:data-prepare}. The recorded trajectories also support our failure analysis by showing how the agent explores the code, interprets available evidence, and develops its final finding. 

\section{Empirical Study Results}
\label{sec:result}


\subsection{RQ1: Effectiveness of Vulnerability Detection}
\subsubsection{Approach}
We evaluate effectiveness at the run level. Each model is evaluated in 3,700 runs, covering the same 37 skills and 100 cases. Each skill is evaluated in 400 runs, covering all four models. For each run, we evaluate three required elements of vulnerability detection: whether the reported vulnerability class matches the ground-truth vulnerability class (CWE correctness), whether the reported function matches an annotated vulnerable function, and whether the reported statement matches an annotated responsible statement (Localization correctness). We then assign the run-level outcome from these three elements. A run is fully correct when all three criteria are satisfied, partial when exactly one or two criteria are satisfied, 
and incorrect when none of the three criteria is satisfied. This rule is applied uniformly across all models, skills, and vulnerability cases.

\subsubsection{Results}
\setlength{\floatsep}{6pt plus 1pt minus 2pt}

\definecolor{softblue}{RGB}{91,128,165}
\definecolor{softorange}{RGB}{213,166,91}
\definecolor{softpurple}{RGB}{145,123,168}
\definecolor{softrose}{RGB}{194,132,132}

\begin{figure}[t!]
\centering
\begin{tikzpicture}

\begin{axis}[
    ybar,
    bar width=8pt,
    width=\linewidth,
    height=3.6cm,
    ymin=0,
    ymax=92,
    ylabel={Percentage (\%)},
    symbolic x coords={
        Fully Correct,
        Partial,
        Incorrect,
        Vulnerabilities Covered
    },
    xtick=data,
    xticklabel style={
        text width=2.2cm,
        align=center,
        font=\small
    },
    ytick={0,20,40,60,80},
    yticklabel style={
        font=\small
    },
    ymajorgrids=true,
    grid style={
        dashed,
        gray!25
    },
    axis line style={
        gray!60
    },
    tick style={
        gray!60
    },
    enlarge x limits=0.10,
    legend style={
        at={(0.5,1.04)},
        anchor=south,
        legend columns=4,
        draw=none,
        font=\small,
        column sep=7pt
    },
    nodes near coords,
    point meta=explicit symbolic
]

\addplot[
    fill=softblue,
    draw=softblue,
    every node near coord/.append style={
        font=\tiny,
        rotate=30,
        anchor=south west,
        xshift=-3pt,
        yshift=-1pt
    }
] coordinates {
    (Fully Correct,5.54) [5.54\%]
    (Partial,18.70) [18.70\%]
    (Incorrect,75.76) [75.76\%]
};

\addplot[
    fill=softorange,
    draw=softorange,
    every node near coord/.append style={
        font=\tiny,
        rotate=30,
        anchor=south west,
        xshift=-3pt,
        yshift=2pt
    }
] coordinates {
    (Fully Correct,12.78) [12.78\%]
    (Partial,15.97) [15.97\%]
    (Incorrect,71.24) [71.24\%]
};

\addplot[
    fill=softpurple,
    draw=softpurple,
    every node near coord/.append style={
        font=\tiny,
        rotate=30,
        anchor=south west,
        xshift=-2pt,
        yshift=-3pt
    }
] coordinates {
    (Fully Correct,5.76) [5.76\%]
    (Partial,15.73) [15.73\%]
    (Incorrect,78.51) [78.51\%]
};

\addplot[
    fill=softrose,
    draw=softrose,
    every node near coord/.append style={
        font=\tiny,
        rotate=30,
        anchor=south west,
        xshift=1pt,
        yshift=3pt
    }
] coordinates {
    (Fully Correct,20.19) [20.19\%]
    (Partial,20.46) [20.46\%]
    (Incorrect,59.35) [59.35\%]
};

\legend{
    Claude Haiku-4.5,
    Claude Sonnet-5,
    DeepSeek-V3.2,
    DeepSeek-V4-Pro
}

\end{axis}

\end{tikzpicture}
\vspace{-0.2cm}
\caption{Model performance and vulnerability coverage. Fully correct, partial, and incorrect results are reported as percentages of the 3,700 runs per model, while vulnerability coverage represents the percentage of the 100 vulnerabilities covered by each model.}
\label{fig:model_performance}
\vspace{-0.2cm}
\end{figure}
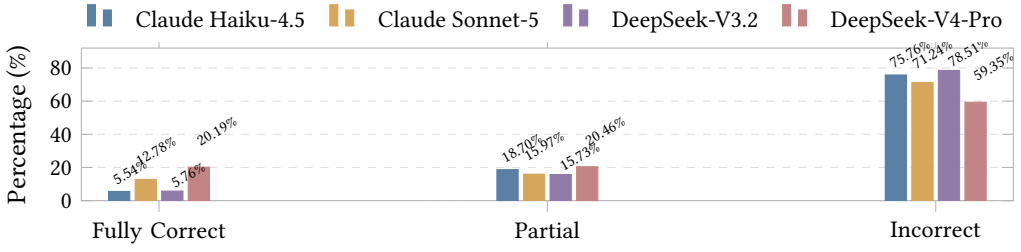
\setlength{\textfloatsep}{5pt plus 1pt minus 2pt}
\begin{figure*}[t]
 \centering
\begin{tikzpicture}
\begin{axis}[
    ybar,
    width=\columnwidth,
    height=3.8cm,
    bar width=4pt,
    ylabel={\footnotesize Matched runs (\%)},
    symbolic x coords={
        Haiku-4.5, Sonnet-5,
        DeepSeek-V3.2, DeepSeek-V4-Pro, All Models
    },
    xtick=data,
    x tick label style={rotate=20,anchor=east,font=\scriptsize},
    ymin=0, ymax=40,
    ytick={0,10,20,30,40},
    enlarge x limits=0.12,
    legend style={
        at={(0.5,1.03)},anchor=south,
        legend columns=4,font=\scriptsize,draw=none
    },
    tick label style={font=\scriptsize},
    nodes near coords,
    nodes near coords style={
        font=\tiny,
        /pgf/number format/fixed,
        /pgf/number format/precision=1
    }
]

\addplot coordinates {
    (Haiku-4.5,20.73)
    (Sonnet-5,26.49)
    (DeepSeek-V3.2,15.59)
    (DeepSeek-V4-Pro,34.76)
    (All Models,24.39)
};

\addplot coordinates {
    (Haiku-4.5,14.73)
    (Sonnet-5,18.73)
    (DeepSeek-V3.2,10.68)
    (DeepSeek-V4-Pro,28.86)
    (All Models,18.25)
};

\addplot coordinates {
    (Haiku-4.5,9.03)
    (Sonnet-5,15.30)
    (DeepSeek-V3.2,11.38)
    (DeepSeek-V4-Pro,27.03)
    (All Models,15.68)
};

\addplot coordinates {
    (Haiku-4.5,5.54)
    (Sonnet-5,12.78)
    (DeepSeek-V3.2,5.76)
    (DeepSeek-V4-Pro,20.19)
    (All Models,11.07)
};

\legend{
    Function,
    Function + Statement,
    CWE,
    Fully Correct
}

\end{axis}
\end{tikzpicture}
\vspace{-0.4cm}
  \caption{Performance across localization and vulnerability identification. 
  For each model, the bars show the percentage of runs that identify the correct function, identify both the correct function and responsible statement, the correct vulnerable class,  and the percentage that satisfy all three criteria (fully correct).}

  \label{fig:localization}
\end{figure*}
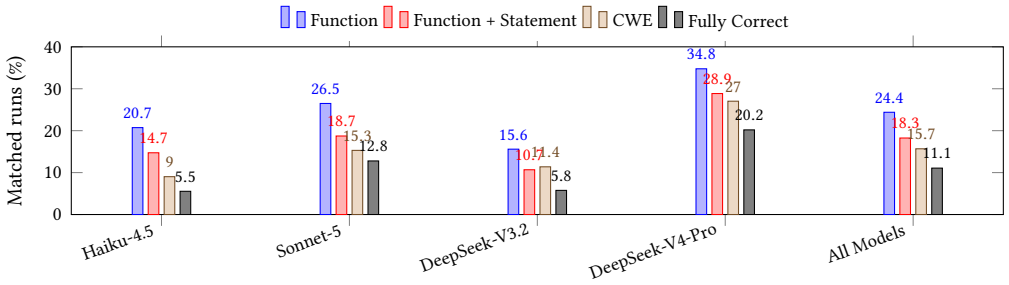

As shown in Fig.~\ref{fig:model_performance}, across the 14,800 runs, 11.07\% satisfy all three correctness conditions. Another 17.72\% of runs identify at least one required element but do not complete the full vulnerability identification, while 71.22\% identify none of the three required elements. DeepSeek-V4-Pro achieves the highest correct rate at 20.19\%, followed by Claude Sonnet-5 at 12.78\%. DeepSeek-V3.2 and Claude Haiku-4.5 achieve 5.76\% and 5.54\%, respectively. 

Figure ~\ref{fig:localization} compares model performance across vulnerability classification, code localization, and their joint correctness. Across all skill-guided model-case runs, 24.4\% locate the correct functions, 18.3\% identify both the correct function and responsible statement, 15.7\% identify the correct CWEs, and produce 11.1\% fully correct results. Localization correctness decreases from 24.4\% at the function level to 18.3\% when statement-level localization is additionally required. This shows that the model may reach the relevant function but still fail to isolate the specific operation that causes the vulnerability. Vulnerability classification presents a separate challenge - only 15.7\% of runs identify the correct CWE. Among the runs with correct function-and-statement localization, 39.4\% still fail to identify the correct vulnerability class. Thus, models may reach the correct code without correctly understanding the type of vulnerability present. 

The performance in vulnerability classification and localization also shows substantial differences across the models. DeepSeek-v4-pro performs strongest across all individual criteria, identifying the correct function in 34.8\% of runs, 28.9\% joint function-statement localization, and the CWE in 27\%, resulting in a fully correct rate of 20.2\%. Claude Sonnet-5 follows with 26.5\%, 18.7\%, and 15.3\%, and 12.8\%, respectively. Claude Haiku-4.5 identifies the correct function relatively often (20.7\%), but shows a much larger decline in CWE identification (9\%), and only 5.5\% of runs are fully correct. DeepSeek-v3.2 has the lowest function-localization rate (15.6\%), while its statement and CWE rates are similar at 10.7\% and 11.4\%, respectively, with a fully correct rate at 5.8\%.

\mybox{
\textbf{Observation O1} The security skills offer partial guidance for Linux kernel vulnerability detection, but reliable end-to-end success remains limited. Across all runs, 24.4\% locate the correct function, 18.3\% identify the responsible statement, and 15.7\% identify the correct CWE, while only 11.1\% correctly identify all three.}

\mybox{\textbf{Observation O2:}
The effectiveness of the same skill-guided analysis varies substantially
across LLMs. Fully correct rates range from 5.54\% to 20.19\%
across the four evaluated models, showing that skill effectiveness also depends on the model executing the procedure.}

\subsection{RQ2: Performance Variation Across Vulnerability Classes and Skill Procedures}
\label{sec:rq2}
\subsubsection{Approach}
We examine performance variation from two complementary perspectives. First, we aggregate runs by CWE class to determine how vulnerability-detection and localization effectiveness varies across the ten weakness types in our benchmark. Second, we compare the 37 representative skills under matched model–vulnerability combinations, giving 400 runs per skill. Because each selected skill comes from a distinct procedure cluster, the evaluation covers a diverse set of vulnerability-analysis workflows. We report effectiveness at the individual-skill level and then inspect the procedures of higher- and lower-performing skills to identify workflow characteristics associated with their observed performance. We focus on three aspects of the procedure: how suspicious code is selected for investigation, how the analysis expands to related state and context, and what evidence is required before a candidate is accepted.

\subsubsection{Results}
\begin{figure*}[t]
\centering

\begin{tikzpicture}

\begin{axis}[
    width=0.80\textwidth,
    height=5.0cm,
    xmin=0,
    xmax=78,
    ymin=0.45,
    ymax=10.70,
    xlabel={Runs matching the criterion (\%)},
    xlabel style={font=\small,yshift=-2pt},
    xtick={0,10,20,30,40,50,60,70},
    xticklabels={0\%,10\%,20\%,30\%,40\%,50\%,60\%,70\%},
    ytick=\empty,
    tick label style={font=\small},
    axis x line*=bottom,
    axis y line=none,
    tick align=outside,
    xmajorgrids=true,
    grid style={
        draw=gray!18,
        line width=0.4pt
    },
    clip=false,
    legend style={
        at={(0.50,1.08)},
        anchor=south,
        draw=none,
        font=\small,
        legend columns=3,
        column sep=1.2em,
        /tikz/every even column/.append style={column sep=0.5cm}
    }
]


\addplot[
    draw=blue!20,
    line width=1pt,
    forget plot
] coordinates {(0.41,1) (1.76,1) (8.72,1)};

\addplot[
    draw=blue!20,
    line width=1pt,
    forget plot
] coordinates {(1.96,2) (9.46,2) (16.01,2)};

\addplot[
    draw=blue!20,
    line width=1pt,
    forget plot
] coordinates {(2.16,3) (10.10,3) (18.70,3)};

\addplot[
    draw=blue!20,
    line width=1pt,
    forget plot
] coordinates {(3.45,4) (8.30,4) (13.20,4)};

\addplot[
    draw=blue!20,
    line width=1pt,
    forget plot
] coordinates {(3.78,5) (11.35,5) (15.61,5)};

\addplot[
    draw=blue!20,
    line width=1pt,
    forget plot
] coordinates {(4.05,6) (9.30,6) (17.30,6)};

\addplot[
    draw=blue!20,
    line width=1pt,
    forget plot
] coordinates {(11.62,7) (13.58,7) (19.53,7)};

\addplot[
    draw=blue!20,
    line width=1pt,
    forget plot
] coordinates {(14.53,8) (19.19,8) (23.65,8)};

\addplot[
    draw=blue!20,
    line width=1pt,
    forget plot
] coordinates {(29.19,9) (37.16,9) (45.07,9)};

\addplot[
    draw=blue!20,
    line width=1pt,
    forget plot
] coordinates {(39.53,10) (62.23,10) (66.28,10)};


\addplot[
    only marks,
    mark=square,
    mark size=3pt,
    blue,
    mark options={
        fill=white,
        draw=blue,
        line width=1pt
    }
] coordinates {
    (8.72,1)
    (16.01,2)
    (18.70,3)
    (13.20,4)
    (15.61,5)
    (17.30,6)
    (19.53,7)
    (23.65,8)
    (45.07,9)
    (66.28,10)
};
\addlegendentry{Function localized}


\addplot[
    only marks,
    mark=diamond,
    mark size=3.4pt,
    orange,
    mark options={
        fill=white,
        draw=orange,
        line width=1pt
    }
] coordinates {
    (1.76,1)
    (9.46,2)
    (10.10,3)
    (8.30,4)
    (11.35,5)
    (9.30,6)
    (13.58,7)
    (19.19,8)
    (37.16,9)
    (62.23,10)
};
\addlegendentry{Function + statement}


\addplot[
    only marks,
    mark=*,
    mark size=3.2pt,
    green!55!black,
    mark options={
        fill=green!55!black,
        draw=green!55!black
    }
] coordinates {
    (0.41,1)
    (1.96,2)
    (2.16,3)
    (3.45,4)
    (3.78,5)
    (4.05,6)
    (11.62,7)
    (14.53,8)
    (29.19,9)
    (39.53,10)
};
\addlegendentry{Fully correct}


\node[
    anchor=west,
    font=\small\bfseries
] at (axis cs:71.4,10.65) {Correct (\%)};


\node[anchor=west,font=\small\bfseries]
    at (axis cs:73,10) {39.53};

\node[anchor=west,font=\small\bfseries]
    at (axis cs:73,9) {29.19};

\node[anchor=west,font=\small\bfseries]
    at (axis cs:73,8) {14.53};

\node[anchor=west,font=\small\bfseries]
    at (axis cs:73,7) {11.62};

\node[anchor=west,font=\small\bfseries]
    at (axis cs:73,6) {4.05};

\node[anchor=west,font=\small\bfseries]
    at (axis cs:73,5) {3.78};

\node[anchor=west,font=\small\bfseries]
    at (axis cs:73,4) {3.45};

\node[anchor=west,font=\small\bfseries]
    at (axis cs:73,3) {2.16};

\node[anchor=west,font=\small\bfseries]
    at (axis cs:73,2) {1.96};

\node[anchor=west,font=\small\bfseries]
    at (axis cs:73,1) {0.41};


\node[anchor=east,font=\small]
    at (axis cs:-1.5,10) {CWE-787};

\node[anchor=east,font=\small]
    at (axis cs:-1.5,9) {CWE-125};

\node[anchor=east,font=\small]
    at (axis cs:-1.5,8) {CWE-476};

\node[anchor=east,font=\small]
    at (axis cs:-1.5,7) {CWE-416};

\node[anchor=east,font=\small]
    at (axis cs:-1.5,6) {CWE-825};

\node[anchor=east,font=\small]
    at (axis cs:-1.5,5) {CWE-401};

\node[anchor=east,font=\small]
    at (axis cs:-1.5,4) {CWE-362};

\node[anchor=east,font=\small]
    at (axis cs:-1.5,3) {CWE-617};

\node[anchor=east,font=\small]
    at (axis cs:-1.5,2) {CWE-908};

\node[anchor=east,font=\small]
    at (axis cs:-1.5,1) {CWE-667};

\end{axis}

\end{tikzpicture}

\caption{ Vulnerability detection and localization performance across CWE classes. Classes are ordered by fully correct rate.
}
\label{fig:cwe-detection-comparison}

\end{figure*}
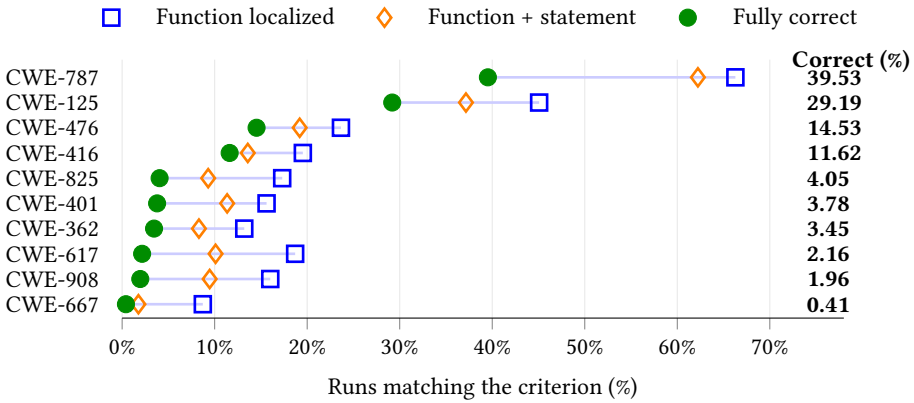
\begin{figure*}[t]
\centering
\begin{tikzpicture}
\begin{axis}[
    xbar,
    width=0.85\textwidth,
    height=9cm,
    xmin=0,
    xmax=19,
    bar width=3pt,
    enlarge y limits=0.06,
    xlabel={Fully correct rate (\%)},
    ylabel={Skill},
    symbolic y coords={ 
        plugin-security-audit,
        sickn33-differential-review,
        abstract-state-analyzer,
        anthropic-vuln-scan,
        posthog-contracts-security,
        mcouthon-security-review,
        security-review-deep,
        ai-devkit-security-review,
        vulnerability-pattern-matcher,
        find-bugs,
        code-vulnerability-review,
        workerd-safety-review,
        agentsmesh-security-auditor,
        source-code-audit,
        cybersecurity-review,
        cortex-bug-hunter,
        secure-code-review,
        bug-hunter-hunter,
        sendaifun-vulnhunter,
        code-security-review,
        commit-security-scan,
        crewrig-security,
        posthog-security-audit,
        ecmwf-security-audit,
        sast-gap-analyzer,
        security-auditor,
        vulnerability-analysis,
        sast-flow-tracer,
        yaklang-code-review,
        audit-design-weaknesses,
        forge-bughunter,
        scrutineer-vuln-scan,
        security-deep-dive,
        memory-api-misuse,
        concurrency-lens,
        vbs-scan-security,
        sc-race-condition
    },
    ytick=data,
    y dir=reverse,
    tick label style={font=\scriptsize},
    label style={font=\small},
    title style={font=\small},
    xmajorgrids,
    grid style={dashed,gray!30},
    axis x line*=bottom,
    axis y line*=left,
    nodes near coords,
    every node near coord/.append style={
        font=\scriptsize,
        anchor=west,
        xshift=2pt
    },
    point meta=x,
    nodes near coords={
        \pgfmathprintnumber[fixed,precision=2]{\pgfplotspointmeta}\%
    },
    clip=false
]
\addplot coordinates {
    (15.75,plugin-security-audit)
    (15.50,sickn33-differential-review)
    (15.00,abstract-state-analyzer)
    (15.00,anthropic-vuln-scan)
     (14.50,mcouthon-security-review)
    (14.50,posthog-contracts-security)
    (14.00,security-review-deep)
    (14.00,ai-devkit-security-review)
    (13.50,vulnerability-pattern-matcher)
    (13.50,find-bugs)
    (12.75,code-vulnerability-review)
    (12.75,workerd-safety-review)
    (12.50,agentsmesh-security-auditor)
    (12.25,source-code-audit)
    (12.25,cybersecurity-review)
    (12.00,cortex-bug-hunter)
    (12.00,secure-code-review)
    (11.50,bug-hunter-hunter)
    (11.25,sendaifun-vulnhunter)
    (11.25,code-security-review)
    (11.25,commit-security-scan)
    (11.00,crewrig-security)
    (11.00,posthog-security-audit)
    (11.00,ecmwf-security-audit)
    (10.75,sast-gap-analyzer)
    (10.50,security-auditor)
    (10.00,vulnerability-analysis)
    (9.75,sast-flow-tracer)
    (9.50,yaklang-code-review)
    (8.75,audit-design-weaknesses)
    (8.25,forge-bughunter)
    (7.25,scrutineer-vuln-scan)
    (6.75,security-deep-dive)
    (5.75,memory-api-misuse)
    (4.75,concurrency-lens)
    (4.50,vbs-scan-security)
    (3.00,sc-race-condition)
};
\end{axis}
\end{tikzpicture}
\caption{Fully correct rates of the 37 evaluated skills, ranked from highest to lowest. Each bar shows the percentage of fully correct runs out of 400 total runs for that skill.}
\label{fig:skill-correct-rate}
\vspace{-0.3cm}
\end{figure*}
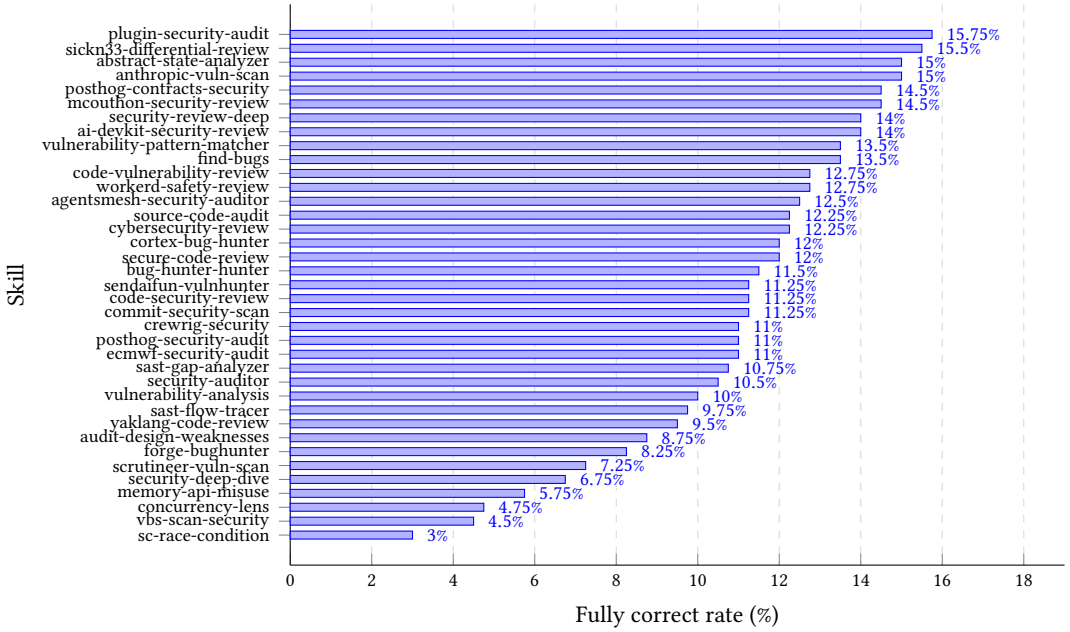

For each CWE class, we aggregate all 1,480 runs and report the number and proportion that satisfy each localization and correctness criterion as shown in Fig. ~\ref{fig:cwe-detection-comparison}. Performance varies substantially across both vulnerability classes and skill-guided workflows. As shown in Fig.~\ref{fig:cwe-detection-comparison}, CWE-787 and CWE-125 achieve the highest correct rates, at 39.53\% and 29.19\%, respectively. CWE-476 and CWE-416 are in a second group, with correct rates of 14.53\% and 11.62\%, respectively. In contrast, CWE-401, CWE-825, CWE-362, CWE-617, CWE-908, and especially CWE-667 remain below 5\%. The localization results help explain these differences. For CWE-787, 66.3\% of runs identify the responsible function, and 62.2\% also identify the responsible statement, demonstrating that once the vulnerable function is found, 93.9\% of these runs also reach the relevant operation. This pattern is consistent with the structure of many bounds violations, where the key evidence is relatively local: an index, length, copy size, or parser offset can be related directly to the size of the accessed object. CWE-667 presents a contrast pattern. Although 8.7\% of runs reach the correct function, only 1.8\% identify the responsible statement, and the full correctness rate is very low. 

To isolate the effect of the skill, we compare the 37 skills while holding both the model and the vulnerability fixed. With four models and 100 vulnerabilities, this gives 400 model-vulnerability combinations. 
As shown in Fig.~\ref{fig:skill-correct-rate}, the fully correct rate ranges from 3\% to 15.75\%, with plugin-security-audit achieving the highest rate at 15.75\%, followed by sickn33-differential-review at 15.5\%, abstract-state-analyzer and anthropic-vuln-scan at 15\%, and posthog-contracts-security at 14.5\%. Although these skills use different procedures, all require the initial hypothesis to undergo an additional semantic or verification step before it becomes a final finding. For example, plugin-security-audit represents a scanner-led hybrid workflow, whereas abstract-state-analyzer and anthropic-vuln-scan represent workflows that separate candidate generation from later verification. Compared with high-performing skills, lower-performing skills more often begin with a narrower analytical lens, such as predefined sinks, resource operations, trust boundaries, or targeted search expressions. These focused workflows can be effective when the chosen lens matches the vulnerability, but they may miss defects whose evidence lies outside that initial search structure. 
\vspace{-0.3cm}
\mybox{
\textbf{Observation O3:} Vulnerability classes differ substantially in how well skill-guided agents can detect them. Bounds-related vulnerabilities are the two highest-performing classes (39.53\%, 29.19\%), while several resource-, state-, initialization-, and synchronization-related classes remain below 5\%. }

\mybox{\textbf{Observation O4:} Across the 37 representative skills, fully correct rates range from 3\% to 15.75\%. Inspection of their procedures shows that several higher-performing representatives more sufficiently combine broad candidate discovery with a subsequent reasoning or verification stage to establish the complete vulnerability evidence chain 
}

\subsection{RQ3: Failure Mechanisms}
\label{sec:rq3}


\subsubsection{Approach}
To understand why skill-guided runs fail, we analyze the unsuccessful execution corpus using the recorded agent trajectories, final findings, evaluation reports, ground-truth vulnerabilities, and the procedures specified by each skill. The trajectories capture which code the agent inspects, what its tool returns, and how the analysis develops before the final response. We develop an LLM-based automatic evaluator due to the large experiment base to conduct a corpus-wide inductive analysis of the unsuccessful runs and identify recurring failure mechanisms. The evaluator is provided with the ground-truth vulnerability, evaluation report, available trajectory evidence, and relevant skill procedure, and it is designed to explain the cause of failure, compare recurring patterns across runs, merge conceptually similar mechanisms, and distinguish failures arising from the agent's reasoning and execution from limitations in the skill instructions themselves. This corpus-wide analysis produces a preliminary failure taxonomy. Following prior work using LLMs for inductive theme and taxonomy generation at scale, we treat the resulting categories as provisional until independently validated by human annotators \cite{liu2026humanllmcollaborativeinductivecoding, jiang2026demystifyingagentskillsworkuntil}. The analysis distinguishes two broad sources of failure: agent-execution failures and skill-level limitations. Agent execution failures capture how the investigation breaks down despite available instructions or evidence, whereas skill-level limitations capture cases where the procedure itself does not adequately support the required investigation or reasoning. This separation is motivated by prior trajectory-analysis research showing that failures emerge through agent execution while their characteristics can also depend on the surrounding agent scaffold, making it important to distinguish execution-level failures from limitations in the guidance provided to the agent~\cite{be6389057634462b87635dfebf6c42f1, zhao2026failureprocessanatomycli}. For each unsuccessful run, the evaluator assigns one primary failure mechanism. The mechanism is attributed either to agent execution or to a skill-level limitation. We validate the taxonomy on a statistically sampled subset of unsuccessful runs. Using a 95\% Confidence Interval and a 5\% margin of error, we sample 374 cases for human validation. We recruit 5 graduate students with a background in software engineering and security, and each annotator independently assigns the applicable failure category based on the given ground truth and the analysis report. Each sampled case was independently labeled by two human annotators. The two annotation sets have 77.8\% inter-rater agreement. Disagreement cases are subsequently adjudicated to obtain the final human-validated labels. Comparing these labels with the original LLM assignments, 80.5\% of the LLM labels were confirmed by human validation.
The final validated taxonomy is used to characterize the recurring reasons that skill-guided agents fail to correctly identify and localize vulnerabilities. 

\subsubsection{Results}
\begin{table*}[t]
\centering
\caption{Failure mechanisms identified from unsuccessful vulnerability-detection runs.}
\label{tab:failure_taxonomy}

\scriptsize
\setlength{\tabcolsep}{3pt}
\renewcommand{\arraystretch}{0.8}

\begin{tabularx}{\textwidth}{
    @{}
    >{\raggedright\arraybackslash}p{3 cm}
    >{\centering\arraybackslash}p{0.7cm}
    >{\raggedright\arraybackslash}X
    >{\raggedleft\arraybackslash}p{1.6cm}
    @{}
}
\toprule
\textbf{Failure Type} &
\textbf{ID} &
\textbf{Category} &
\makecell[r]{\textbf{Share of All}\\\textbf{Failure Labels}} \\
\midrule

\multirow{7}{1.7cm}{\raggedright\textbf{Agent\\Execution\\Failure}}
& \textbf{M1}
& Relevant evidence not analyzed
& 29.3\% \\

& \textbf{M2}
& Required state relation not constructed
& 11.7\% \\

& \textbf{M3}
& Incorrect contract or code interpretation
& 7.6\% \\

& \textbf{M4}
& Unsupported or contradicted argument
& 4.9\% \\

& \textbf{M5}
& Candidate closed without resolving the decisive question
& 4.2\% \\

& \textbf{M6}
& Investigation action not carried out effectively
& 0.9\% \\

& \textbf{M7}
& Finding lost or altered between analysis and output
& 1.3\% \\

\midrule

\multirow{7}{1.7cm}{\raggedright\textbf{Skill-Level\\Limitation}}
& \textbf{S1}
& Task and scope adaptation gap
& 3.8\% \\

& \textbf{S2}
& Kernel reasoning procedure gap
& 15.6\% \\

& \textbf{S3}
& Investigation and coverage-control gap
& 15.8\% \\

& \textbf{S4}
& Standalone execution and dependency gap
& 1.9\% \\

& \textbf{S5}
& Evidence verification and completion gap
& 2.3\% \\

& \textbf{S6}
& Candidate retention and triage gap
& 0.8\% \\

& \textbf{S7}
& Finding contract gap
& 0.05\% \\

\bottomrule
\end{tabularx}
\end{table*}

\begin{figure*}[t!]
    \centering

    \begin{subfigure}[t]{0.49\textwidth}
        \centering
        \includegraphics[
            width=\linewidth,
            height=6cm,
            keepaspectratio
        ]{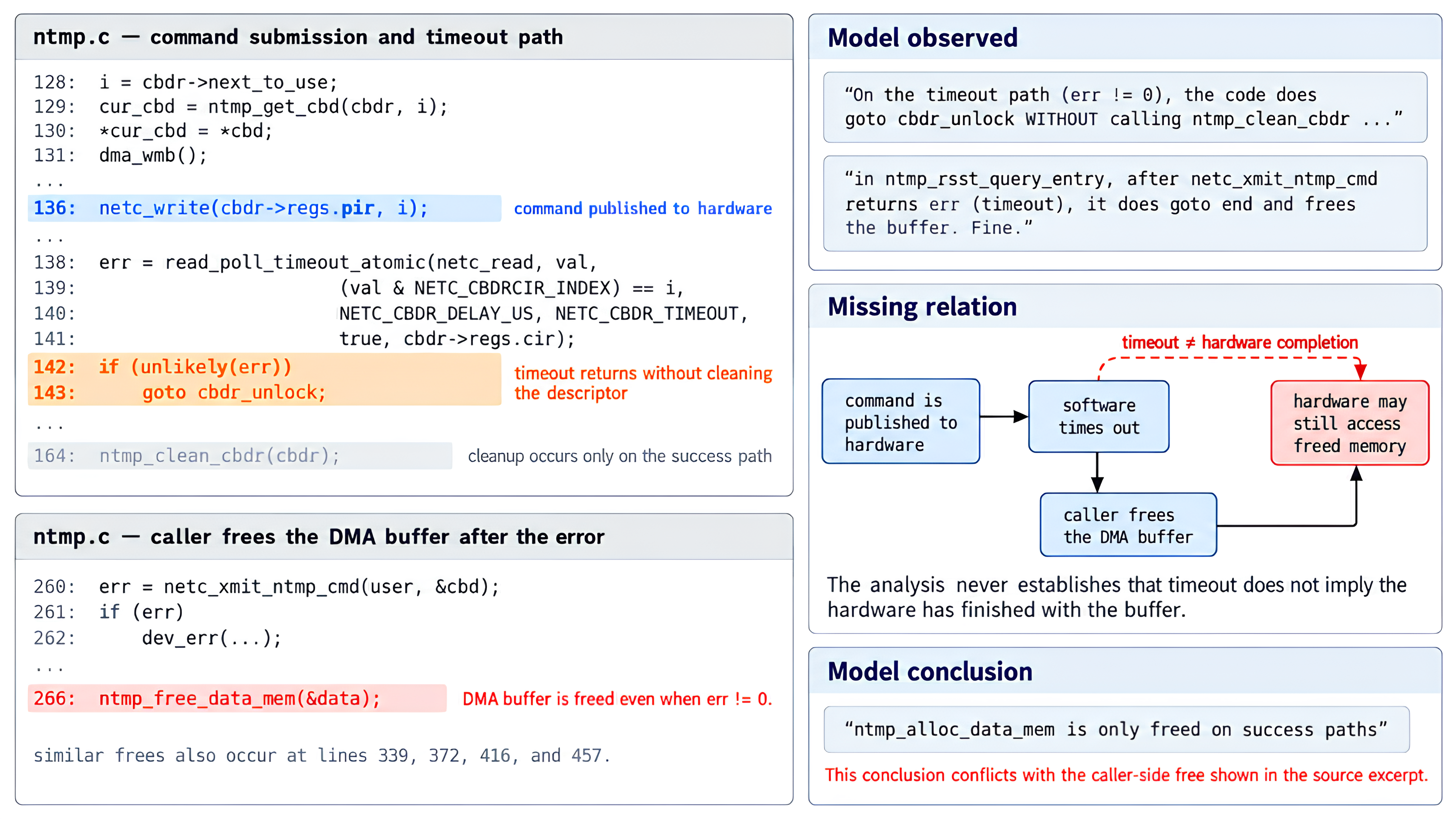}
        \caption{Example of M2 -- the agent identifies the timeout path and the buffer free, but does not connect them through the hardware ownership relation, and therefore misses the use-after-free vulnerability.}
        \label{fig:m2_example}
    \end{subfigure}
    \hfill
    \begin{subfigure}[t]{0.49\textwidth}
        \centering
        \includegraphics[
            width=\linewidth,
            height=6cm,
            keepaspectratio
        ]{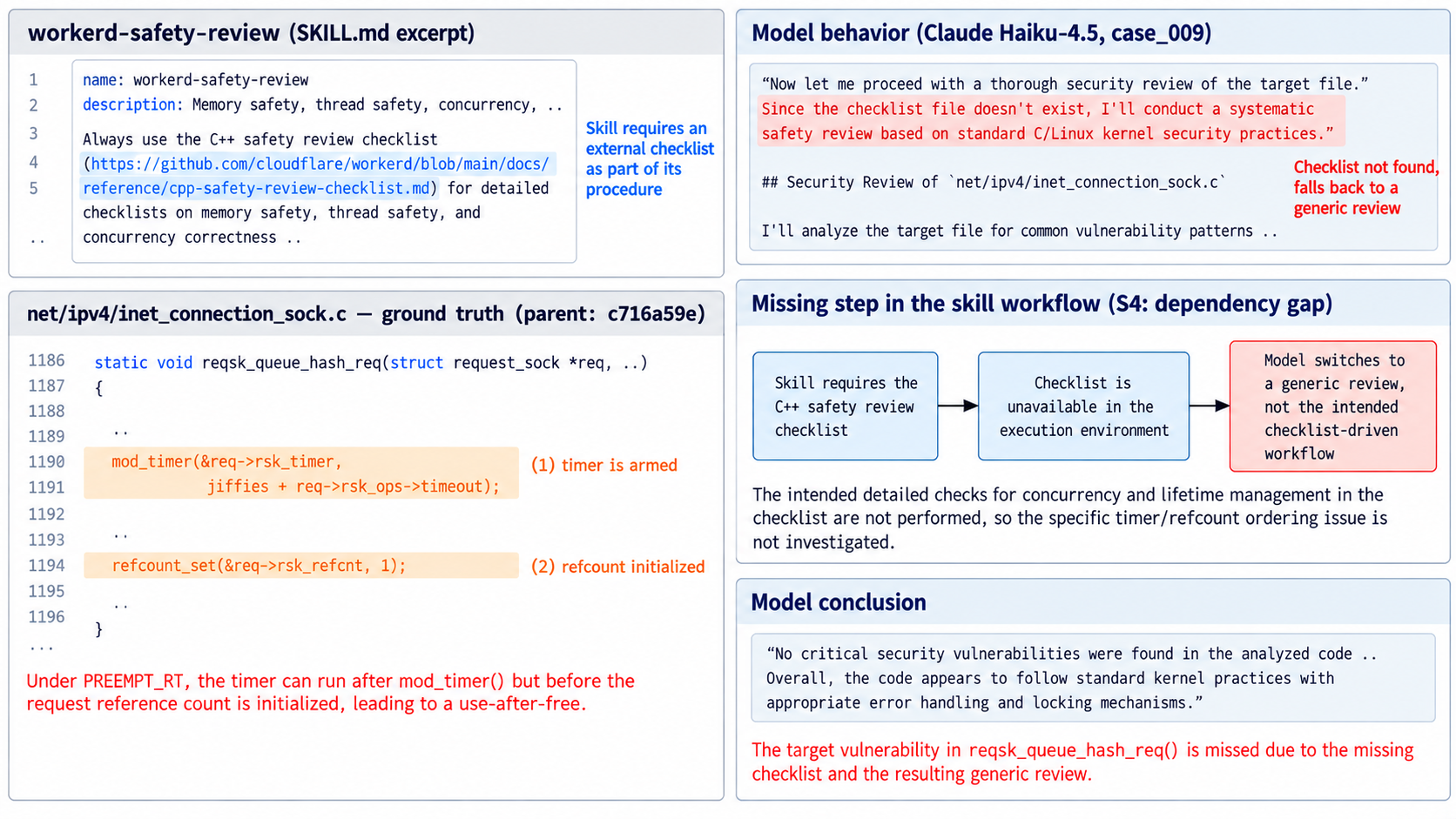}
        \caption{Example of S4 -- the run replaces the intended checklist procedure with a generic review, and the target timer/refcount ordering analysis is never performed.}
        \label{fig:s4_example}
    \end{subfigure}

    \caption{Examples of agent-execution and skill-level failure mechanisms.}
    \label{fig:failure_examples}
    
\end{figure*}

Table~\ref{tab:failure_taxonomy} summarizes the validated failure taxonomy. Across the two sources of primary failure in agent execution and skill limitation, the observed failure mechanisms span three parts of the vulnerability-detection process: evidence acquisition and investigation, construction and verification of the vulnerability argument, and candidate reporting. 
\vspace{-0.3cm}
\paragraph{Evidence Acquisition and Investigation} Some failures occur before a complete vulnerability argument can be formed. At the agent level, the category of \textbf{relevant evidence not analyzed} captures cases where relevant code is available but is not meaningfully analyzed, while the category of \textbf{investigating action not carried out effectively} captures ineffective execution of an investigation step after the missing evidence has already been recognized. The corresponding skill-level limitations include, for example, \textbf{an investigation and coverage-control gap}, where the procedure does not adequately specify which callers, callbacks, or initialization must be covered, and a \textbf{standalone execution and dependency gap}, where the workflow depends on unavailable files, tools, or prior-stage artifacts without providing an effective alternative (an example is shown in Fig.~\ref{fig:s4_example}). 
\paragraph{Construction and verification of the vulnerability argument.} A second group of failures occurs after relevant operations have already been found. \textbf{Required state relation not constructed} captures cases where the agent fails to construct the ownership, temporal, synchronization, initialization, bounds, or error-state relation that makes the operations unsafe (an example is shown in Fig.~\ref{fig:m2_example}). \textbf{Incorrect contract or code interpretation} instead reflects an incorrect interpretation of a concrete API, helper, guard, state, lock, or execution context, while \textbf{Unsupported or contradicted argument} captures conclusions that are unsupported by or inconsistent with the available evidence. On the skill side, \textbf{task and scope adaptation gap} reflects a review scope that does not adapt to the reasoning  required by the target; the \textbf{kernel reasoning procedure gap} captures the absence of a concrete kernel-specific reasoning  procedure, and the \textbf{evidence verification and completion gap} reflects insufficient requirements for verifying a finding or safety 
conclusion before completing the analysis.
\paragraph{Candidate retention and reporting.} Failures can also occur after a relevant candidate has already been identified. \textbf{Candidate closed without resolving the decisive question} captures premature closure, where the agent rejects or stops investigating a candidate before resolving the condition needed to confirm or refute it. \textbf{Finding lost or altered between analysis and output} captures cases where a supported finding is subsequently omitted, altered, or loses the mechanism or locations required in the final report. The corresponding skill-level mechanisms are \textbf{candidate retention and triage gap}, where triage rules allow unresolved candidates to be discarded too early, and the \textbf{finding and handoff contract gap}, where the output or stage-handoff contract does not preserve the complete finding.
\mybox{
\textbf{Observation O5:} Failure mechanisms arise from both agent execution and limitations in the skill procedure. The most frequent individual mechanism is relevant evidence not analyzed is 29.3\%, followed by investigation and coverage-control gaps (15.8\%).
}

\mybox{\textbf{Observation O6:} Failures can occur throughout the lifetime of a vulnerability candidate: relevant evidence may be missed, the vulnerability relation may remain unresolved, a candidate may be accepted or rejected without sufficient verification, or a supported finding may be lost during reporting. Across different failed runs, the primary mechanisms may arise either from agent execution or from limitations in the skill procedure.}
\section{From Failure Analysis to Skill Self-Improvement}
\label{sec: skill-sif}

The empirical study in Section~\ref{sec:result} establishes how security skills perform and where skill-guided executions break down. We therefore extend the study from empirical characterization to skill improvement, aiming to address whether a security skill can be progressively strengthened through its own execution experience. To this end, we develop {\tool}, an iterative self-improving framework that diagnoses unsuccessful executions, extracts failure lessons, and progressively refines security skills based on execution feedback. 

\subsection{{\tool} Framework Overview}

\begin{figure*}[t!]
  \centering  \includegraphics[width=\textwidth]{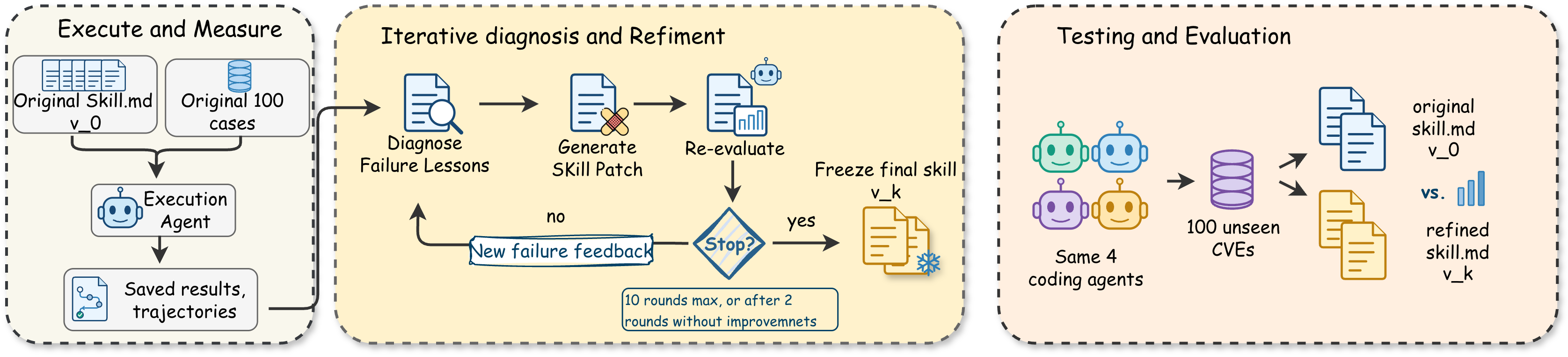}
  \caption{Iterative Skill Self-Improving Framework} 
\label{fig:sif}
\vspace{-0.2cm}

\end{figure*}

Figure ~\ref{fig:sif} presents {\tool}, a failure-guided evolutionary refinement framework that progressively evolves a security skill from its own execution experience. Starting from an initial SKILL.md, {\tool} repeatedly executes the skill on vulnerable programs, evaluates each run against the ground truth, and analyzes unsuccessful trajectories to identify the reasoning or procedural breakdowns that prevented a fully correct finding. The case-level failures are aggregated into recurring failure lessons, which capture missing investigation steps, reasoning relations, or verification requirements. The refinement model then translates these lessons into a targeted procedural mutation of the current skill, producing the next skill generation. The revised skill is re-executed over the refinement set, allowing each generation to be evaluated through both newly solved cases and regressions. Through this execute-diagnose-evolve-validate cycle, {\tool} converts repeated vulnerability-detection failures into persistent procedural knowledge and progressively strengthens the skill's ability to guide security analysis.


\subsection{Skill Refinement Setup}
\label{sec:refinement-setup}

Given a skill, we use {\tool} to iteratively evolve it. Each skill is evolved independently using the 100 Linux kernel vulnerabilities from our empirical study as the refinement set. We use gpt-5.6-luna as the refinement model to provide a consistent evolution environment across skills and separate the refinement process from the four models used later for evaluation. For each skill revision, the current skill is executed on all refinement cases and evaluated using the same CWE, function, statement, and fully correct criteria as the empirical study. Failure trajectories are diagnosed and aggregated into failure lessons, which drive the next procedural mutation of SKILL.md. Every revised generation is then re-executed and re-evaluated in the next round. Due to budget constraints, the evolution process is limited to at most 10 rounds and terminates early if the best fully correct score does not improve for two consecutive rounds. The highest-performing generation is then frozen and used in the testing evaluation.

\section{Evaluation of {\tool}}
\label{sec: skill-sif-performance}


We now evaluate whether the refined skills improve vulnerability detection on previously unseen cases and how refinement changes the failure mechanisms that remain. We therefore investigate the following two research questions: 
\begin{itemize}[leftmargin=10pt]
    \item \textbf{RQ4: How effectively does {\tool} improve security skills for detecting and localizing the vulnerabilities?}
    \item \textbf{RQ5: How does skill refinement change the failure mechanisms observed in skill-guided vulnerability detection?}
\end{itemize}

\subsection{Evaluation Setup}
\label{sec:sif-eval-setup}
\subsubsection{Experimented Skills}
We select the five top-performing skills identified in the empirical study~\ref{sec:rq2} to evaluate the performance of {\tool}: plugin-security-audit, sickn33-differential-review, abstract-state-analyzer, anthropic-vuln-scan, and mcouthon-security-review.

\subsubsection{Evaluation Dataset}
To evaluate whether the refined skills generalize beyond the cases used during refinement, we construct a new set of 100 previously unseen Linux kernel vulnerabilities. These cases are unseen during the refinement loop. For each of the five selected skills, we evaluate both the original and refined versions on the same testing cases using the same four underlying LLMs from the empirical study, which provides matched baseline vs.\ refined comparisons for the same skill, model, and vulnerability case. We use the same evaluation measures as in the empirical study: correct vulnerability class (CWE), correct vulnerable function, function-and-statement localization, and fully correct detection. 


\subsection{RQ4: Effectiveness of {\tool}}
\label{sec:rq4}

\begin{table*}[t]
\centering
\caption{Fully correct refinement cases by refinement round. Each cell gives the score out of 100, followed by newly correct cases ($+G$) and regressions ($-L$) relative to the preceding round. Bold scores identify the frozen revisions used for testing; \texttt{-} denotes a round not executed.}
\label{tab:sif-rounds}
\footnotesize
\setlength{\tabcolsep}{1pt}
\renewcommand{\arraystretch}{0.6}
\begin{tabular}{@{}lccccc@{}}
\toprule
\diagbox{\textbf{Round}}{\textbf{Skills}} & \makecell{\textbf{security-audit}} & \makecell{\textbf{differential-review}} & \makecell{\textbf{abstract-state-analyzer}} & \makecell{\textbf{vuln-scan}} & \makecell{\textbf{security-review}} \\
\midrule
0 & $4$ & $7$ & $5$ & $6$ & $2$ \\
1 & $13\;(+11,-2)$ & $12\;(+8,-3)$ & $11\;(+8,-2)$ & $\mathbf{11}\;(+6,-1)$ & $\mathbf{14}\;(+13,-1)$ \\
2 & $14\;(+5,-4)$ & $12\;(+4,-4)$ & $\mathbf{12}\;(+4,-3)$ & $7\;(+1,-5)$ & $11\;(+3,-6)$ \\
3 & $13\;(+3,-4)$ & $13\;(+5,-4)$ & $9\;(+1,-4)$ & $10\;(+6,-3)$ & $9\;(+4,-6)$ \\
4 & $14\;(+4,-3)$ & $14\;(+5,-4)$ & $9\;(+2,-2)$ & $11\;(+7,-6)$ & - \\
5 & $14\;(+5,-5)$ & $14\;(+4,-4)$ & - & $9\;(+4,-6)$ & - \\
6 & $\mathbf{14}\;(+5,-5)$ & $16\;(+7,-5)$ & - & $8\;(+3,-4)$ & - \\
7 & - & $12\;(+5,-9)$ & - & - & - \\
8 & - & $\mathbf{17}\;(+7,-2)$ & - & - & - \\
9 & - & $17\;(+3,-3)$ & - & - & - \\
10 & - & $14\;(+6,-9)$ & - & - & - \\
\midrule
\textbf{Gain over round 0} & $+10$ & $+10$ & $+7$ & $+5$ & $+12$ \\
\bottomrule
\end{tabular}
\end{table*}
\definecolor{cwehead}{HTML}{4E79A7}
\definecolor{cwesub}{HTML}{6D93BD}
\definecolor{cweshade}{HTML}{EAF1FB}

\definecolor{funhead}{HTML}{59A14F}
\definecolor{funsub}{HTML}{79B770}
\definecolor{funshade}{HTML}{ECF6EA}

\definecolor{fshead}{HTML}{F2B134}
\definecolor{fssub}{HTML}{F5C35A}
\definecolor{fsshade}{HTML}{FFF6E5}

\definecolor{fullhead}{HTML}{E15759}
\definecolor{fullsub}{HTML}{EA7A7B}
\definecolor{fullshade}{HTML}{FDECEC}

\begin{table*}[t]
\centering
\caption{Correctness rate (\%) on 100 new cases per skill and model.
Each pair reports baseline (B) $\rightarrow$ refined (R).
CWE: CWE correctness; Function: function localization;
F+S: joint function-and-statement localization;
Full: joint correctness. $\Delta$ denotes the change from baseline
to refined in percentage points.}
\label{tab:sif-test}

\scriptsize
\setlength{\tabcolsep}{3pt}
\renewcommand{\arraystretch}{0.8}

\begin{tabular}{@{}p{2.8cm}l
>{\columncolor{cweshade}}c >{\columncolor{cweshade}}c
>{\columncolor{funshade}}c >{\columncolor{funshade}}c
>{\columncolor{fsshade}}c >{\columncolor{fsshade}}c
>{\columncolor{fullshade}}c >{\columncolor{fullshade}}c@{}}
\toprule
\multirow{2}{*}{\textbf{Skill}} &
\multirow{2}{*}{\textbf{LLM}} &
\multicolumn{2}{>{\columncolor{cwehead}}c}{\color{white}\textbf{CWE}} &
\multicolumn{2}{>{\columncolor{funhead}}c}{\color{white}\textbf{Function}} &
\multicolumn{2}{>{\columncolor{fshead}}c}{\color{white}\textbf{F+S}} &
\multicolumn{2}{>{\columncolor{fullhead}}c}{\color{white}\textbf{Full}} \\
\cmidrule(lr){3-4}
\cmidrule(lr){5-6}
\cmidrule(lr){7-8}
\cmidrule(lr){9-10}
&
&
\cellcolor{cwesub}\color{white}\textbf{B$\rightarrow$R} &
\cellcolor{cwesub}\color{white}\textbf{$\Delta$} &
\cellcolor{funsub}\color{white}\textbf{B$\rightarrow$R} &
\cellcolor{funsub}\color{white}\textbf{$\Delta$} &
\cellcolor{fssub}\color{white}\textbf{B$\rightarrow$R} &
\cellcolor{fssub}\color{white}\textbf{$\Delta$} &
\cellcolor{fullsub}\color{white}\textbf{B$\rightarrow$R} &
\cellcolor{fullsub}\color{white}\textbf{$\Delta$} \\
\midrule

plugin-security-audit
& Sonnet-5
& $10\rightarrow18$ & $+8$
& $20\rightarrow30$ & $+10$
& $19\rightarrow26$ & $+7$
& $9\rightarrow14$ & $+5$ \\

& Haiku-4.5
& $7\rightarrow5$ & $-2$
& $7\rightarrow12$ & $+5$
& $6\rightarrow9$ & $+3$
& $5\rightarrow4$ & $-1$ \\

& DeepSeek-V4-Pro
& $4\rightarrow21$ & $+17$
& $5\rightarrow34$ & $+29$
& $4\rightarrow29$ & $+25$
& $2\rightarrow16$ & $+14$ \\

& DeepSeek-V3.2
& $9\rightarrow12$ & $+3$
& $10\rightarrow20$ & $+10$
& $7\rightarrow16$ & $+9$
& $5\rightarrow7$ & $+2$ \\

\midrule

sickn33-differential-review
& Sonnet-5
& $14\rightarrow27$ & $+13$
& $17\rightarrow36$ & $+19$
& $17\rightarrow35$ & $+18$
& $14\rightarrow24$ & $+10$ \\

& Haiku-4.5
& $2\rightarrow9$ & $+7$
& $8\rightarrow9$ & $+1$
& $5\rightarrow8$ & $+3$
& $2\rightarrow8$ & $+6$ \\

& DeepSeek-V4-Pro
& $6\rightarrow19$ & $+13$
& $7\rightarrow26$ & $+19$
& $7\rightarrow21$ & $+14$
& $5\rightarrow11$ & $+6$ \\

& DeepSeek-V3.2
& $7\rightarrow14$ & $+7$
& $11\rightarrow21$ & $+10$
& $7\rightarrow17$ & $+10$
& $4\rightarrow9$ & $+5$ \\

\midrule

abstract-state-analyzer
& Sonnet-5
& $10\rightarrow19$ & $+9$
& $12\rightarrow31$ & $+19$
& $10\rightarrow28$ & $+18$
& $6\rightarrow16$ & $+10$ \\

& Haiku-4.5
& $4\rightarrow4$ & $0$
& $7\rightarrow8$ & $+1$
& $6\rightarrow7$ & $+1$
& $3\rightarrow4$ & $+1$ \\

& DeepSeek-V4-Pro
& $13\rightarrow16$ & $+3$
& $17\rightarrow14$ & $-3$
& $15\rightarrow12$ & $-3$
& $13\rightarrow11$ & $-2$ \\

& DeepSeek-V3.2
& $7\rightarrow11$ & $+4$
& $10\rightarrow16$ & $+6$
& $7\rightarrow13$ & $+6$
& $5\rightarrow7$ & $+2$ \\

\midrule

anthropic-vuln-scan
& Sonnet-5
& $9\rightarrow23$ & $+14$
& $15\rightarrow28$ & $+13$
& $13\rightarrow25$ & $+12$
& $9\rightarrow21$ & $+12$ \\

& Haiku-4.5
& $5\rightarrow6$ & $+1$
& $11\rightarrow9$ & $-2$
& $10\rightarrow7$ & $-3$
& $4\rightarrow4$ & $0$ \\

& DeepSeek-V4-Pro
& $15\rightarrow20$ & $+5$
& $16\rightarrow23$ & $+7$
& $12\rightarrow21$ & $+9$
& $8\rightarrow17$ & $+9$ \\

& DeepSeek-V3.2
& $6\rightarrow13$ & $+7$
& $14\rightarrow10$ & $-4$
& $10\rightarrow9$ & $-1$
& $4\rightarrow8$ & $+4$ \\

\midrule

mcouthon-security-review
& Sonnet-5
& $6\rightarrow25$ & $+19$
& $7\rightarrow25$ & $+18$
& $6\rightarrow25$ & $+19$
& $5\rightarrow17$ & $+12$ \\

& Haiku-4.5
& $4\rightarrow6$ & $+2$
& $6\rightarrow6$ & $0$
& $3\rightarrow5$ & $+2$
& $3\rightarrow2$ & $-1$ \\

& DeepSeek-V4-Pro
& $14\rightarrow16$ & $+2$
& $17\rightarrow18$ & $+1$
& $14\rightarrow16$ & $+2$
& $12\rightarrow12$ & $0$ \\

& DeepSeek-V3.2
& $5\rightarrow16$ & $+11$
& $8\rightarrow18$ & $+10$
& $6\rightarrow16$ & $+10$
& $3\rightarrow12$ & $+9$ \\

\bottomrule
\end{tabular}
\end{table*}

Table~\ref{tab:sif-rounds} presents the round-by-round results for the refinement sequences from which the frozen skills were selected. The five skills follow different convergence patterns: plugin-security-audit is refined for six rounds, sickn33-differential-review for ten, abstract-state-analyzer for four, anthropic-vuln-scan for six, and mcouthon-security-review for three. The selected versions come from rounds 6, 8, 2, 1, and 1, respectively. Across the five skills, fully correct detections increase from 24 to 68 out of 500 skill-case runs, corresponding to an improvement from 4.8\% to 13.6\% (+8.8 percentage points). All five skills improve over their original versions, ranging from 5 to 12 percentage points. Most of the improvement occurs early in refinement. After the first revision, the five skills increase from 24 to 61 fully correct runs. After the first revision, the aggregate number of fully correct runs increases from 24 to 61, a net gain of 37. This represents 84.1\% of the overall net increase of 44 runs achieved by the selected skill versions. Later rounds still provide meaningful gains for some skills. For example, sickn33-differential-review increases from 12 fully correct cases after its first revision to 17 at round 8. In contrast, anthropic-vuln-scan and mcouthon-security-review reach their highest fully correct scores after the first revision, with later modifications failing to produce further net gains. 

We next evaluate whether the improvements obtained during refinement transfer to the 100 new collected vulnerabilities. Table~\ref{tab:sif-test} compares the original and refined versions of each of the five skills using the same four underlying LLMs from the empirical study. Aggregated across all 2,000 skill-model-case runs, fully correct detection improves from 6.05\% to 11.20\% (+5.15 percentage points). The gains also appear consistently across the intermediate evaluation criteria: CWE correctness increases from 7.85\% to 15\%, function localization from 11.25\% to 19.7\%, and joint function-statement localization from 9.2\% to 17.25\%. Therefore, the refined skills improve both the ability to reach the responsible code and the ability to correctly detect the vulnerability. The improvement also transfers across all four underlying LLMs. Sonnet-5 shows the largest aggregate gain in fully correct detection, increasing from 8.6\% to 18.4\% (+9.8pp) across the five skills. DeepSeek-v4-pro improves from 8\% to 13.4\% (+5.4pp), DeepSeek-v-3.2 from 4.2\% to 8.6\% (+4.4pp), and Haiku-4.5 from 3.4\% to 4.4\% (+1.0pp). The same pattern is visible in localization and CWE correctness: for example, Sonnet-5 increases function localization from 14.2\% to 30\%, while DeepSeek-V4-Pro increases from 12.4\% to 23\%. At the skill level, all five refined skills achieve a higher aggregate fully correct rate across the four models. plugin-security-audit increases from 5.25\% to 10.25\%, sickn33-differential-review from 6.25\% to 13\%, abstract-state-analyzer from 6.75\% to 9.5\%, anthropic-vuln-scan from 6.25\% to 12.5\%, and mcouthon-security-review from 5.75\% to 10.75\%. At the individual skill--model level, 15 of the 20 combinations improve in fully correct detection, two remain unchanged, and three decline. The largest gain occurs for plugin-security-audit with DeepSeek-V4-Pro, which increases from 2\% to 16\% (+14pp). 

\mybox{\textbf{Observation O7:} {\tool} improves the five selected skills during refinement, and the gains transfer to unseen vulnerabilities. Across the four LLMs on the testing set, fully correct detection increases from 6.05\% to 11.2\%, while CWE correctness, function localization, and joint function-and-statement localization also improve.  }

\mybox{\textbf{Observation O8:} The magnitude of improvements varies across LLMs, with gains ranging from +1 to +9.8 percentage points, which shows that the refined procedures generalize across models while their effectiveness remains dependent on how each agent executes the guidance. }

\subsection{RQ5: Changes in Failure Mechanisms}
\input{tabs/category_change}

Figure~\ref{fig:failure-shift} compares the failure mechanisms observed under the baseline and refined skills on the test cases. Refinement reduces the total number of unsuccessful runs from 1,879 to 1,776. More importantly, the composition of the remaining failures changes substantially. Under the baseline skills, skill-level limitations (S1-S7) account for 8.67\% of unsuccessful runs. After refinement, this share falls to only 0.06\%. S1-S4 disappear entirely, S5 decreases from 0.53\% to 0.06\%, and S6-S7 remain absent in both settings. Thus, the procedural limitations identified in the baseline skills are almost absent from runs guided by the refined skills. The refined skills produce 103 fewer unsuccessful runs, while nearly eliminating the skill-level limitations captured by our taxonomy. Failures that remain are overwhelmingly associated with agent execution failure mechanisms, particularly M1 and M4. This pattern is consistent with the role of skill refinement: revising SKILL.md can improve the analysis procedure and remove deficiencies in the guidance provided to the agent, but it cannot directly eliminate limitations in how the underlying model retrieves evidence, constructs relations, interprets code, or carries out the prescribed procedure. 

\mybox{\textbf{Observation O9}: Skill refinement substantially reduces skill-level limitations in testing executions. Skill-level mechanisms account for 8.67\% of baseline failures but only 0.06\% after refinement, while the total number of unsuccessful runs decreases from 1,879 to 1,776. }


\section{Discussion}
\label{sec:discussion}
\begin{table*}[t]
\centering
\caption{Design implications derived from the empirical observations.}
\label{tab:skill-design-implications}
\small
\setlength{\tabcolsep}{3pt}
\renewcommand{\arraystretch}{0.7}

\begin{tabularx}{\textwidth}{@{}X p{2.5cm}@{}}
\toprule
\textbf{Security-Skill Design Takeaway} &
\textbf{Observations} \\
\midrule

\textbf{Provide broad and structured investigation coverage.}
Require analysis of related callers, callbacks, initialization, error,
cleanup, and teardown paths rather than relying on a narrow initial search.
& O3, O4, O5 \\

\textbf{Require explicit security-relation reasoning and verification.}
Make ownership, lifetime, bounds, synchronization, initialization, temporal,
and error-state relations explicit and verify them before concluding.
& O1, O3, O4, O5, O7 \\

\textbf{Define candidate-resolution and reporting rules.}
Keep unresolved candidates active until decisive evidence is checked, and
preserve the mechanism, class, location, and supporting evidence in the final finding.
& O1, O6 \\

\textbf{Use explicit, model-robust procedures.}
Specify analysis steps, decision conditions, evidence requirements, and
stopping rules to reduce reliance on model-specific implicit reasoning.
& O2, O5, O6, O8 \\

\bottomrule
\end{tabularx}
\end{table*}
\vspace{-0.1cm}

Our observations suggest four practical considerations for designing future vulnerability-detection skills, summarized in Table~\ref{tab:skill-design-implications}.

\textbf{Ensure broad and structured investigation coverage.}
Higher-performing skills generally provide sufficient coverage to generate candidate vulnerabilities and then expand the analysis beyond the initial search lens. RQ3 further shows that missing relevant evidence and insufficient coverage control are major sources of failure. Skills should therefore specify
which related paths and contexts need to be examined, such as callers, callbacks, initialization, error paths, cleanup, and teardown, while allowing the investigation to expand when the initial search is insufficient.

\textbf{Require explicit security-relation reasoning and verification.}
Our results show that vulnerabilities involving lifetime, resources,
synchronization, initialization, and execution state are substantially harder
than locally observable bounds violations. The failure analysis further shows that agents often reach relevant operations without constructing the relation that makes them unsafe. Skills should explicitly guide reasoning about ownership, lifetime, bounds, synchronization, temporal order, and error state, and require these relations to be verified against the source before reaching a conclusion.

\textbf{Define clear candidate-resolution and reporting rules.}
Reaching the relevant code does not necessarily produce a complete vulnerability finding. Candidates may be closed before the decisive condition is resolved, and supported findings may lose important information in the final output. Skills should therefore define the evidence required to confirm or reject a candidate and require the final finding to preserve the vulnerability mechanism, class, responsible function and statement, and supporting evidence.

\textbf{Write skills as explicit, model-robust procedures.}
The effectiveness of the same procedural guidance varies across models,
and the improvements obtained from refined skills also differ by model. Important reasoning steps should therefore not be left implicit. Investigation
steps, decision conditions, evidence requirements, and stopping rules should be stated operationally so that the intended workflow depends less on model-specific reasoning behavior.

Finally, RQ5 illustrates the boundary of what can be addressed through the skill layer. After refinement, the observed skill-level limitations nearly disappear from the remaining unsuccessful runs, while failures associated with agent execution remain. This suggests that stronger procedures can reduce gaps in
the supplied guidance, but cannot eliminate failures arising from how the
underlying agent executes that guidance.
\section{Threats To Validity}
\label{sec:threats}
Our evaluation defines a fully correct detection as jointly identifying the
ground-truth CWE, vulnerable function, and responsible statement. This strict criterion may understate partially useful analyses, although partial outcomes are reported separately. Ground truth is derived from fixing patches and allows multiple valid functions and statements when a vulnerability spans multiple locations. The failure taxonomy also involves interpretive judgment, which we
mitigate through independent human validation on 374 sampled failures followed by adjudication of disagreements. Skill collection, filtering, and procedure clustering may also affect the final evaluation set. We reduce this risk through cross-model screening, manual resolution of screening disagreements, and manual
review of procedure clusters. However, because one representative skill is
selected from each eligible cluster, the measured performance applies to the
selected skill rather than every skill sharing a similar workflow. SKILL-SIF also uses a single refinement model, so different refinement agents may produce different revisions. 
The external validity of our findings is limited by the study scope. We focus
on Linux-kernel vulnerabilities, ten CWE classes, a file-guided analysis setting, four underlying LLMs, and, for refinement, the five highest-performing skills. The results may therefore differ for other software ecosystems, vulnerability classes, agent models, or less constrained auditing environments. We partially mitigate this limitation by evaluating procedurally diverse skills, models from two families, vulnerabilities spanning ten weakness classes, and
refined skills on a separate set of previously unseen vulnerabilities.

\section{Conclusion}
\label{sec:conclusion}

In this paper, we conducted a large-scale empirical study of procedural security skills for agent-based vulnerability detection, evaluating 37 public skills across four LLMs and 100 real Linux kernel vulnerabilities. Our results show that existing skills provide useful but limited guidance: only 11.1\% of 14,800 runs correctly identified the vulnerability class, responsible function, and statement, with substantial variation across vulnerability classes and skill workflows. Our failure analysis further reveals recurring breakdowns in evidence acquisition, state-relation reasoning, verification, candidate retention, and reporting. Building on these findings, we introduced {\tool}, an iterative self-improving framework that learns from failed executions to refine security skills. Evaluation on 100 unseen vulnerabilities shows that the refined skills improve fully correct detection by up to 8$\times$, demonstrating that execution feedback can systematically improve security skills and generalize to unseen vulnerabilities.


\bibliographystyle{ACM-Reference-Format}
\bibliography{ref}

\end{document}